\documentclass[12pt]{article}
\usepackage{amscd}
\usepackage{amsfonts}
\usepackage{CJK}
\usepackage{amssymb}
\usepackage{amsmath}
\usepackage{color}
\usepackage{cite}

\newcommand\btd{\raise 2pt \hbox{$\hat\bigtriangledown$}\hskip 1.5pt}
\newcommand\bt{\raise 2pt \hbox{$\bigtriangledown$}\hskip 1.5pt}
\def\no{\nonumber}

\begin{document}
\title{Approximate nonlinear self-adjointness and approximate conservation laws}
\author{Zhi-Yong Zhang \footnote{E-mail: zhiyong-2008@163.com; Tel:+86 010 88803103} 
\\\small~College of Sciences, North China University of
Technology, Beijing 100144, P.R. China}
\date{}
\maketitle

\noindent{\bf Abstract:} In this paper, approximate nonlinear
self-adjointness for perturbed PDEs is introduced and its properties
are studied. Consequently, approximate conservation laws which
cannot be obtained by approximate Noether theorem, are constructed
by means of the method. As an application, a class of perturbed
nonlinear wave equations is considered to illustrate the
effectiveness.


\section{Introduction}
When the German mathematician Emmy Noether proved her theorem, she
established a connection between symmetries  and conservation laws
of differential equations, provided that the equations under
consideration are obtained from a variational principle, i.e., they
are Euler-Lagrange equations \cite{noether,21}.  In order to invoke
this powerful theorem, one requires a Lagrangian of the underlying
differential equations which make it cast as an Euler-Lagrange
system \cite{nhi1}. It is well known that there is no Lagrangian for
scalar evolution equations, such as the classical heat equation, the
Burgers equations, etc. \cite{ander-1984}. In consequence, one
cannot associate conservation laws with their symmetries via
Noether's theorem.  Thus, given a system without a Lagrangian
formulation, one needs a corresponding algorithm to find
conservation laws of the system. In \cite{blu11,blu21}, Anco and
Bluman proposed a direct construction formula of local conservation
laws for partial differential equations (PDEs) expressed in a
standard Cauchy-Kovalevskaya form. Kara and Mahomed presented a
partial Noether approach, which is efficient for Euler-Lagrange type
equations \cite{kara-2006}.

Recently, the general concept of nonlinear self-adjointness
\cite{ib-2011,nhi},  which includes strict self-adjointness
\cite{nh-2006,nh-2007}, quasi self-adjointness \cite{ib-2007}  and
weak self-adjointness \cite{gan-2011} stated earlier, was introduced
to construct conservation laws associated with symmetries of
differential equations. The main idea of the method traced back to
\cite{ba-1931,ca-1988} and followed in \cite{Olver} (see Exercise
5.37). The method introduced a formal Lagrangian of the system
consisted of the governing equations together with their adjoint
equations and then utilized the conservation law theorem in
\cite{nh-2007} to construct local and nonlocal conservation laws of
the PDEs under study.

Owing to the fast development of nonlinear self-adjointness and its
subclasses, many important physical PDEs have been studied
successfully
\cite{ib-2011,Fs-2012,fre-2013,tt-2012,gan-2012,nhi-2011}. For
example,  the required conditions of self-adjointness and quasi
self-adjointness for a class of third order PDEs were presented in
\cite{tt-2012}. Gandarias and Bruz$\acute{\mbox{o}}$n considered
conservation laws of a forced KdV equation via weak self-adjointness
\cite{gan-2012}. Nonlinear self-adjointness of a generalized
fifth-order KdV equation was studied in \cite{Fs-2012}. The authors
in \cite{nhi-2011} showed that a (2+1)-dimensional generalized
Burgers equation written as a system of two dependent variables was
quasi self-adjoint. Further examples can be found in
\cite{ib-2011,fre-2013} and references therein.

Another vital achievement in the past several decades is the
emergence of approximate symmetry, which aims to deal with the
differential equations with a small parameter possessing few exact
symmetries or none at all and even if exist, the small parameter
also disturbs symmetry group properties of the unperturbed equation.
Consequently, two reasonably well-known approaches originated from
Baikov et al. \cite{ba-1991} and Fushchich and Shtelen
\cite{fs-1989} arose, which employed standard perturbation
techniques about the symmetry operator and dependent variables to
obtain approximate symmetry respectively. In \cite{pak,ron}, these
two methods were applied to three nonlinear PDEs which showed that
the second method was superior to the first one. Systematic methods
for obtaining both exact symmetries and first-order approximate
symmetries for ordinary differential equations (ODEs) were available
in \cite{ib-1999}.

In the meantime, the theory of approximate conservation laws
associated with perturbed differential equations was introduced with
regard to approximate Noether symmetries, i.e., symmetries
associated with a Lagrangian of the perturbed differential equations
\cite{kara-2002,ah}. In \cite{ah-1999}, the authors studied how to
construct approximate conservation laws for perturbed PDEs via
approximate generalized symmetries. In \cite{ag-2006}, a basis of
approximate conservation laws for perturbed PDEs was discussed.
Johnpillai et.al \cite{ah-2009} showed how to construct approximate
conservation laws of approximate Euler-type equations via
approximate Noether type symmetry operators associated with partial
Lagrangians.  Quite recently, the concept of self-adjointness was
extended to tackle perturbed PDEs and successfully applied to study
two examples to obtain approximate conservation laws \cite{ib-2011}.
However, the study of approximate conservation law is still a major
object for both mathematician and physicist and should be further
developed.

The purpose of the paper is to perform a further study of the
properties and applications of approximate nonlinear
self-adjointness for perturbed PDEs. The outline of the paper is
arranged as follows. In Section 2, some related basic notions and
principles are reviewed and the definition of approximate nonlinear
self-adjointness and its properties are given. In Section 3, the
method is applied to a class of perturbed nonlinear wave equations
and approximate conservation laws are constructed. The last section
contains a conclusion of our results.
\section{Main results}
We first recall some basic notions and principles associated with
approximate symmetry and nonlinear self-adjointness in the first two
subsections, and then give main results about approximate nonlinear
self-adjointness for perturbed PDEs in the last subsection.

Consider a system of $m$ PDEs with $r$th-order
\begin{eqnarray}\label{perturbnew}
E_{\alpha} = E_{\alpha}^0(x,u,u_{(1)},\cdots,u_{(r)})+ \epsilon
E_{\alpha} ^1(x,u,u_{(1)},\cdots,u_{(r)})=0,
\end{eqnarray}
where $\epsilon$ is a small parameter, $\alpha=1,\dots,m$,
$x=(x^1,\dots,x^n),u=(u^1,\dots,u^m)$, $u^{\sigma}_i=\partial
u^{\sigma}/\partial x^i$, $u^{\sigma}_{ij}=\partial^2
u^{\sigma}/\partial x^i\partial x^j$,\dots,  and $u_{(i)}$ denotes
the collection of all $i$th-order partial derivatives of $u$ with
respect to $x$, e.g., $u_{(i)}=\{u^{\sigma}_i\}$ with
$\sigma=1,\dots,m$.  Note that we will use these symbols and the
summation convention for repeated indices throughout the paper if no
special notations are added.

System (\ref{perturbnew}) is called perturbed PDEs while the system
which do not contain the perturbed term $\epsilon E_{\alpha}
^1(x,u,u_{(1)},\cdots,u_{(r)})$, i.e.,
\begin{eqnarray}\label{perturb}
E_{\alpha}^0(x,u,u_{(1)},\cdots,u_{(r)})=0,
\end{eqnarray}
is called unperturbed PDEs.

The classical method for obtaining exact (Lie point) symmetries
admitted by PDEs (\ref{perturb}) is to find  a one-parameter local
transformation group
\begin{eqnarray}\label{group-1}
&&\no (x^i)^*=x^i+\epsilon \,\xi^i(x,u)+O(\epsilon^2),\\
&&(u^{\sigma})^*=u^{\sigma}+\epsilon
\,\eta^{\sigma}(x,u)+O(\epsilon^2),
\end{eqnarray}
which leaves system (\ref{perturb}) invariant. Lie's method requires
that the infinitesimal generator of transformation (\ref{group-1}),
i.e., $X=\xi^i(x,u)\partial_{x^i}
+\eta^{\sigma}(x,u)\partial_{u^{\sigma}}$,  satisfies Lie's
infinitesimal criterion
\begin{equation}\label{deter-1}
\text{pr}^{(k)}X(E_{\alpha}^0)=0, ~~~\mbox{when}~ E_{\alpha}^0=0,
\end{equation}
where $\text{pr}^{(k)}X$ stands for $k$-order prolongation of $X$
calculated by the well-known prolongation formulae
\cite{blu1,Olver}. The infinitesimal, namely $\xi^i,\eta^{\sigma}$,
can be found from  an over-determined linear system generated by
condition (\ref{deter-1}). We refer to references \cite{blu1,Olver}
for details.
\subsection{Approximate symmetry}
Up until now, there exist two methods to obtain approximate symmetry
of perturbed PDEs.

Firstly, we introduce the method originated from Fushchich and
Shtelen. This method employs a perturbation of dependent variables,
that is, expanding the dependent variable with respect to the small
parameter $\epsilon$ yields
\begin{eqnarray}\label{tran}
&& u^\sigma =\sum^{\infty}_{k=0}\epsilon^k u^{\sigma_k},\qquad
0<\epsilon \ll 1,
\end{eqnarray}
where $u^{\sigma_k}$ are new introduced dependent variables, after
inserting expansion (\ref{tran}) into system (\ref{perturbnew}),
then approximate symmetry is defined as the exact symmetry of the
system corresponding to each order in the small parameter
$\epsilon$. We refer to \cite{fs-1989} for further details.

The second approach, initiated by Baikov et al., is no perturbation
of the dependant variables but a perturbation of the symmetry
generator \cite{ba-1991}.

A first-order approximate symmetry of system (\ref{perturbnew}),
with the infinitesimal operator form $X = X_0 +\epsilon X_1$,  is
obtained by solving for $X_1$ in
\begin{eqnarray}\label{second}
&& X_1(E_{\alpha}^0)_{\mid_{E_{\alpha}^0=0}}+H=0,
\end{eqnarray}
where the auxiliary function $H$ is obtained by
\begin{eqnarray}\label{auxiliary}
\no H=\frac{1}{\epsilon}X_0(E_{\alpha})_{\mid_{E_{\alpha}=0}}.
\end{eqnarray}
$X_0$ is an exact symmetry of unperturbed PDEs $E_{\alpha}^0=0$. The
notation $|_{\Delta=0}$, hereinafter, means evaluation on the
solution manifold of $\Delta=0$.

We formulate the second method as follows.

\textbf{Definition 1.} (Approximate symmetry \cite{ba-1991}) A
first-order approximate symmetry with infinitesimal operator $X =
X_0 +\epsilon X_1$ leaves system (\ref{perturbnew}) approximate
invariant if $X_0$ is an exact symmetry of unperturbed PDEs
$E_{\alpha}^0=0$ and $X_1$ is defined by (\ref{second}).

 The first method by Fushchich and Shtelen uses only
standard Lie algorithm and can be implemented in computer algebra
system, then this approximate symmetry approach may readily be
extended to determine infinite-dimensional and other types
approximate symmetries \cite{zhang1}. As for the second method,
since the dependent variables are not expanded in a perturbation
series, approximate solutions obtained by using a first-order
approximate generator may contain higher-order terms \cite{pak}.
\subsection{Nonlinear self-adjointness}
In this subsection, we briefly recall the main idea of nonlinear
self-adjointness of PDEs in order to induce approximate nonlinear
self-adjointness.

Let $\mathcal {L}$ be the formal Lagrangian of system
(\ref{perturb}) given by
\begin{eqnarray}\label{lagrangian}
&& \mathcal {L} = v^{\beta}E_{\beta}^0(x,u,u_{(1)},\cdots,u_{(r)}),
\end{eqnarray}
then the adjoint equations of system (\ref{perturb}) are defined by
\begin{eqnarray}\label{adequation}
(E_{\alpha}^0)^{\ast}(x,u,v,u_{(1)},v_{(1)},\cdots,u_{(r)},v_{(r)})=\frac{\delta\mathcal
{L}}{\delta u^{\sigma}}=0,
\end{eqnarray}
where $v=(v^{1},\dots,v^m)$ and $v_{(i)}$ represents all $i$th-order
derivatives of $v$ with respect to $x$, $\delta/\delta u^{\sigma}$
is the variational derivative written as
\begin{eqnarray}
\no\frac{\delta}{\delta u^{\sigma}}=\frac{\partial}{\partial
u^{\sigma}}+\sum_{s=1}^{\infty}(-1)^sD_{i_1}\dots
D_{i_s}\frac{\partial}{\partial u^{\sigma}_{i_1\dots i_s}},
\end{eqnarray}
where, hereinafter, $D_i$ denotes the total derivative operators
with respect to $x^i$. For example, a dependent variable $w=w(y,z)$
with $y=x^1,z=x^2$, one has
\begin{eqnarray}
&&\no D_y = \frac{\partial}{\partial y} +w_y
\frac{\partial}{\partial w} +w_{yy} \frac{\partial}{\partial w_y}
+w_{yz}\frac{\partial}{\partial w_z} +\cdots, etc.
\end{eqnarray}

In what follows, we recall the definition of nonlinear
self-adjointness of differential equations.

\textbf{Definition 2.} (Nonlinear self-adjointness \cite{ib-2011})
The system (\ref{perturb}) is said to be nonlinearly self-adjoint if
the adjoint system (\ref{adequation}) is satisfied for all solutions
$u$ of system (\ref{perturb}) upon a substitution
\begin{eqnarray}\label{sub}
v^{\sigma}=\varphi^{\sigma}(x,u),~~~~\sigma=1,\dots,m,
\end{eqnarray}
such that $\varphi(x,u)=(\varphi^{1},\dots,\varphi^m)\neq 0$.

The substitution (\ref{sub}) satisfying $\varphi(x,u)\neq 0$ solves
the adjoint equations (\ref{adequation}) for all solutions of system
(\ref{perturb}), which can be regarded as an equivalent definition
of nonlinear self-adjointness \cite{ib-2011,nhi}. Definition 2 is
also equivalent to the following identities holding for the
undetermined coefficients $\lambda_{\alpha}^{\beta}$
\begin{eqnarray}\label{iden-non}
(E_{\alpha}^0)^{\ast}(x,u,v,u_{(1)},v_{(1)},\cdots,u_{(r)},v_{(r)})_{|_{\{v=\varphi,\dots,v_{(r)}=\varphi_{(r)}\}}}
=\lambda_{\alpha}^{\beta}E_{\beta}^0(x,u,u_{(1)},\cdots,u_{(r)}),
\end{eqnarray}
which is applicable in the computations. Hereinafter,
$\varphi_{(i)}$, similar as $v_{(i)}$ and $u_{(i)}$, stands for all
$i$th-order partial derivatives of $\varphi$ with respect to $x$.

\textbf{Remarks 1.}

1. If the substitution (\ref{sub}) becomes $v^{\sigma}=u^{\sigma}$,
then system (\ref{perturb}) is called strict self-adjointness. If
$v^{\sigma}=\varphi^{\sigma}(u)$ is independent of $x$, then it is
named by quasi self-adjointness. If
$v^{\sigma}=\varphi^{\sigma}(x,u)$ involving all $x$ and $u$, then
it is called weak self-adjointness.

2. The substitution  (\ref{sub}) can also be extended to the case
$v^{\sigma}=\varphi^{\sigma}(x,u,u_{(i)})$, which embraces the
derivatives of $u$ and is called differential substitution.

Obviously, the concept of quasi self-adjointness and weak
self-adjointness generalize strict self-adjointness. Next, we
consider an example about quasi self-adjointness while the readers
can find the examples for weak self-adjointness in \cite{gan-2011}.
Consider a nonlinear PDE studied in \cite{nhi}
\begin{equation}\label{pde}
u_t-u^2u_{xx}=0,
\end{equation}
which describes the nonlinear heat conduction in solid hydrogen
\cite{nh-2006}. Let the formal Lagrangian $\mathcal
{L}=v(u_t-u^2u_{xx})$, then by means of (\ref{adequation}), its
adjoint equation is
\begin{equation}
\no \frac{\delta\mathcal {L}}{\delta u}=v_t+4uvu_{xx}+u^2
v_{xx}+4uu_xv_x+2vu^2_x=0,
\end{equation}
which is identical to Eq.(\ref{pde}) by the substitution $v=u^{-2}$,
not by $v=u$. It means that Eq.(\ref{pde}) is quasi self-adjoint but
not strictly self-adjoint.

The following theorem will be used to construct conservation laws
 for both unperturbed and perturbed cases \cite{nh-2007}.

\textbf{Theorem 1. }  Any infinitesimal symmetry (Local or nonlocal)
\begin{eqnarray}
&&\no X=\xi^i(x,u,u_{(1)},\dots)\frac{\partial}{\partial
x^i}+\eta^{\sigma}(x,u,u_{(1)},\dots)\frac{\partial}{\partial
u^{\sigma}}
\end{eqnarray}
of system (\ref{perturb}) leads to a conservation law $D_i(C^i)=0$
constructed by the formula
\begin{eqnarray}\label{formula}
&&\no C^i=\xi^i\mathcal {L}+W^{\sigma}\Big[\frac{\partial \mathcal
{L}}{\partial u_i^{\sigma}}-D_j(\frac{\partial \mathcal
{L}}{\partial u_{ij}^{\sigma}})+D_jD_k(\frac{\partial \mathcal
{L}}{\partial u_{ijk}^{\sigma}})-\dots\Big]\\&&\hspace{1cm}
+D_j(W^{\sigma}) \Big[\frac{\partial \mathcal {L}}{\partial
u_{ij}^{\sigma}}-D_k(\frac{\partial \mathcal {L}}{\partial
u_{ijk}^{\sigma}})+\dots\Big]+D_jD_k(W^{\sigma})\Big[\frac{\partial
\mathcal {L}}{\partial u_{ijk}^{\sigma}}-\dots\Big]+\dots,
\end{eqnarray}
where $W^{\sigma}=\eta^{\sigma}-\xi^ju_j^{\sigma}$ and $\mathcal
{L}$ is the formal Lagrangian. In applying the formula, the formal
Lagrangian  $\mathcal {L}$ should be written in the symmetric form
with respect to all mixed derivatives
$u_{ij}^{\sigma},u_{ijk}^{\sigma},\dots$.

Generally speaking, the term $\xi^i\mathcal {L}$ with $\mathcal {L}$
in the form (\ref{lagrangian}) can be omitted because it vanishes
identically on the solution manifold of the studying PDEs.

In particular, a first-order approximate conserved vector
$C=(C^1,\dots,C^n)$ of system (\ref{perturbnew}) satisfies
$$D_i(C^i)=O(\epsilon^2)$$ for all solutions of $E_\alpha=0$.
\subsection{Approximate nonlinear self-adjointness}

This subsection will concentrate on the study of approximate
nonlinear self-adjointness of perturbed PDEs. The formal Lagrangian
$\mathcal {\widetilde{L}}$ of perturbed system (\ref{perturbnew}) is
given by
\begin{eqnarray}\label{lagrangian1}
\no \mathcal {\widetilde{L}} =
v^{\beta}[E_{\beta}^0(x,u,u_{(1)},\cdots,u_{(r)})+ \epsilon\,
E_{\beta}^1(x,u,u_{(1)},\cdots,u_{(r)})],
\end{eqnarray}
then the adjoint equations of system (\ref{perturbnew}) are written
as
\begin{eqnarray}\label{adequationper}
E_{\alpha}^{\ast}(x,u,v,u_{(1)},v_{(1)},\dots,u_{(r)},v_{(r)})=\frac{\delta\mathcal
{\widetilde{L}}}{\delta u^{\sigma}}=0.
\end{eqnarray}

\textbf{Definition 3.} (Approximate nonlinear self-adjointness) The
perturbed system (\ref{perturbnew}) is called approximate nonlinear
self-adjointness if the adjoint system (\ref{adequationper}) is
approximate satisfied for all solutions $u$ of system
(\ref{perturbnew}) upon a substitution
\begin{eqnarray}\label{sub1}
v^{\sigma}=\varphi^{\sigma}(x,u)+\epsilon\phi^{\sigma}(x,u),~~~~\sigma=1,\dots,m,
\end{eqnarray}
such that not all $\varphi^{\sigma}$ and $\phi^{\sigma}$ are
identically equal to zero.

It should be mentioned that Definition 3 extends the results for
unperturbed PDEs. It means that, if regarding $\epsilon$ as a usual
parameter and replacing
$\varphi^{\sigma}(x,u)+\epsilon\phi^{\sigma}(x,u)$ in the right side
of (\ref{sub1}) by a new function $\psi^{\sigma}(x,u)$, then
Definition 3 is equivalent to Definition 2 for unperturbed case.
Moreover, Definition 3 also extends the results  regarding perturbed
ODEs \cite{kara-2002} to nonlinear PDEs, where the authors
considered the Lagrangian maintaining the order of the perturbed
parameter as stipulated by the given ODEs with Noether's theorem.

Furthermore, some necessary remarks should be demonstrated about
approximate nonlinear self-adjointness.

\textbf{Remarks 2.}

Denote $\varphi(x,u)=(\varphi^{1},\dots,\varphi^m)$ and
$\phi(x,u)=(\phi^{1},\dots,\phi^m)$. $\phi_{(i)}$ stands for the
same meaning as $\varphi_{(i)}$.

1. The required condition in Definition 3 means that the adjoint
equations of system (\ref{perturbnew}) work out
\begin{eqnarray}\label{forper}
&& \no
\hspace{-0.5cm}E_{\alpha}^{\ast}(x,u,v,u_{(1)},v_{(1)},\dots,u_{(r)},v_{(r)})_{|_{\{v=\varphi+\epsilon
\phi,\dots,v_{(r)}=\varphi_{(r)}+\epsilon
\phi_{(r)}\}}}\\
&&
\hspace{0.5cm}-\left[(\lambda_\alpha^\beta+\epsilon\mu_\alpha^\beta)
E^0_{\beta}(x,u,u_{(1)},\dots,u_{(r)})+\epsilon \lambda_\alpha^\beta
E^1_{\beta}(x,u,u_{(1)},\dots,u_{(r)})\right]=O(\epsilon^2),
\end{eqnarray}
with undetermined parameters $\lambda_\alpha^\beta$ and
$\mu_\alpha^\beta$. Equality (\ref{forper}) provides a computable
formula to discriminate approximate nonlinear self-adjointness for
perturbed PDEs.

 2. Similarly, as the unperturbed
case, we can define approximate strict self-adjointness with
$v=u+\epsilon u$ (or $u$ or $\epsilon u$), approximate quasi
self-adjointness with $v=\varphi(u)+\epsilon\phi(u)$ (or
$\varphi(u)$ or $\epsilon\phi(u)$) and approximate weak
self-adjointness with $v=\varphi(x,u)+\epsilon\phi(x,u)$ (or
$\varphi(x,u)$ or $\epsilon\phi(x,u)$) containing all $x$ and $u$.
Approximate differential substitution also holds if $v$ contains the
derivatives of $u$.

3. If the substitution (\ref{sub1}) does not exist for system
(\ref{perturbnew}), i.e., system (\ref{perturbnew}) is not
approximately nonlinearly self-adjoint, then the resulting conserved
vectors will be nonlocal in the sense that they involve the
introduced variable $v$ connected with the physical variable $u$ via
adjoint equations (\ref{adequationper}).


In what follows, we present some properties of approximate nonlinear
self-adjointness.

\textbf{Theorem 2.} If adjoint system (\ref{adequation}) exists
solutions in the form $v^{\sigma}=\epsilon f^{\sigma}(x,u)$ with
some functions $f^{\sigma}(x,u)$, then system (\ref{perturbnew}) is
approximately nonlinearly self-adjoint.

\emph{Proof.} Observe that $\mathcal {\widetilde{L}} =
v^{\beta}E_{\beta}^0+ \epsilon\,v^{\beta} E_{\beta}^1$. If the
substitution for approximate nonlinear self-adjointness of system
(\ref{perturbnew}) is in the form $v=\epsilon \phi(x,u)$, then by
equivalent equality (\ref{forper}) of Definition 3, $\mathcal
{\widetilde{L}}$ is simplified to $\mathcal
{\widehat{L}}=v^{\beta}E_{\beta}^0$ because the second part
$\epsilon\,v^{\beta} E_{\beta}^1$ is second order of $\epsilon$.
Thus in this case, the adjoint equations of system
(\ref{perturbnew}) become $\delta\mathcal {\widehat{L}}/\delta
u^{\sigma}=\delta(v^{\beta}E_{\beta}^0)/\delta u^{\sigma}=0$, which
has the same form as the adjoint equation of system (\ref{perturb}),
so the solutions $v^{\sigma}=\epsilon f^{\sigma}(x,u)$ of adjoint
system (\ref{adequation}) also satisfy $\delta\mathcal
{\widehat{L}}/\delta u^{\sigma}=0$. It means that
$v^{\sigma}=\epsilon f^{\sigma}(x,u)$ is just the required
substitution which make system (\ref{perturbnew}) to be approximate
nonlinear self-adjointness. This proves the result. $\hfill
\blacksquare$

For instance, consider a perturbed nonlinear wave equation
$F=u_{tt}-u_{xx}+\epsilon uu_t=0$ with formal Lagrangian $\mathcal
{\widetilde{L}}=v(u_{tt}-u_{xx}+\epsilon uu_t)$, the adjoint
equation is $F^{\ast}=\delta\mathcal {\widetilde{L}}/\delta u=
v_{tt}-v_{xx}-\epsilon uv_t=0$. The adjoint equation of unperturbed
equation $u_{tt}-u_{xx}=0$ is $v_{tt}-v_{xx}=0$ which has solution
in the form $v=\epsilon (c_1xt+c_2t+c_3x+c_4)$, where not all
arbitrary constants $c_i\,(i=1,\dots,4)$ are zero. Then by Theorem
2, this solution makes $F^{\ast}=-\epsilon^2u(c_1x+c_2)$, thus
equation $F=0$ is approximately nonlinearly self-adjoint.

In particular, for linear perturbed PDEs, we have the following
results.

\textbf{Corollary.} Any system of linear perturbed PDEs is
approximately nonlinearly self-adjoint.

The corollary is a parallel result as unperturbed case
\cite{ib-2011} and can be shown with almost parallel method, thus we
take an example to demonstrate it. Consider the perturbed linear
wave equation $u_{tt}-u_{xx}+\epsilon u_t=0$ whose formal Lagrangian
is $\mathcal {\widetilde{L}}=v(u_{tt}-u_{xx}+\epsilon u_t)$, then
its adjoint equation is $v_{tt}-v_{xx}-\epsilon v_t=0$ which is
independent of $u$, thus any nontrivial solution is a substitution
to make $u_{tt}-u_{xx}+\epsilon u_t=0$ to be approximate nonlinear
self-adjointness.

For the case of one dependent variable of system (\ref{perturbnew}),
namely $u$, we have the following results.

\textbf{Theorem 3.} Eq.(\ref{perturbnew}) is approximately
nonlinearly self-adjoint if and only if it becomes approximately
strictly self-adjoint after multiplied by an appropriate multiplier
$\mu(x,u)+\epsilon\nu(x,u)$.

\emph{Proof.} Suppose Eq.(\ref{perturbnew}) with one dependent
variable written by $E_1=E_1^0+\epsilon E_1^1=0$ is approximately
nonlinearly self-adjoint, for the substitution $v=\varphi+\epsilon
\phi$, one has
\begin{eqnarray}
\no\frac{\delta (v E_1)}{\delta u}_{\mid_{v=\varphi+\epsilon
\phi}}=(\lambda_0+\epsilon\lambda_1)E_1
\end{eqnarray}
which is equivalent to the following system after separating it with
respect to $\epsilon$
\begin{eqnarray}\label{proof2}
\frac{\delta (\varphi E_1^0)}{\delta u}= \lambda_0 E_1^0,
~~~~\frac{\delta (\varphi E_1^1+\phi E_1^0)}{\delta u}= \lambda_1
E_1^0+\lambda_0 E_1^1.
\end{eqnarray}
Note that hereinafter in all equalities we neglect the terms of order $O(\epsilon^2)$.

On the other hand, the conditions for approximate strict
self-adjointness and the variational derivative associated with
equations $E_1=0$ yield
\begin{eqnarray}\label{proof3}
\frac{\delta [\omega (\mu(x,u)+\epsilon\nu(x,u)) E_1]}{\delta
u}_{\mid_{\omega=u+\epsilon u}}=
(\widetilde{\lambda}_0+\epsilon\widetilde{\lambda}_1)\, E_1.
\end{eqnarray}

Inserting dependent variable $\omega=\omega_0+\epsilon\omega_1$ into
(\ref{proof3}) and splitting it with different order of $\epsilon$,
for $\epsilon^0$, we have
\begin{eqnarray}\label{proof11}
&& \no\frac{\delta (\omega_0 \mu E^0_1)}{\delta u}=\omega_0
\frac{\partial \mu}{\partial u} E^0_1+\mu \omega_0\frac{\partial
(E^0_1)}{\partial u}-D_{i}[\mu \omega_0\frac{\partial
(E^0_1)}{\partial u_i}]+D_iD_j[\mu \omega_0\frac{\partial (
E^0_1)}{\partial u_{ij}}]+\dots\\\no&&\hspace{1.9cm}=\omega_0
\frac{\partial \mu}{\partial u} E^0_1+\frac{\delta (\varphi
E^0_1)}{\delta u}\\&&\hspace{1.9cm}=\widetilde{\lambda}_0 E_1^0,
\end{eqnarray}
where, in Eq.(\ref{proof11}), we regard $\omega_0$ as a dependent
variable  in the first equality while in the second equality,
$\varphi=\omega_0\,\mu(x,u)$ is taken as a new dependent variable
instead of $\omega_0$ to obtain the second term. With the condition
$\omega=u+\epsilon u$, one has
\begin{eqnarray}\label{proof4}
&& \frac{\delta (\omega_0 \mu E^0_1)}{\delta
u}_{\mid_{\omega_0=u}}=u \frac{\partial \mu}{\partial u}
E^0_1+\frac{\delta (\varphi E^0_1)}{\delta
u}_{\mid_{\omega_0=u}}=\widetilde{\lambda}_0 E_1^0.
\end{eqnarray}

Similarly, for $\epsilon^1$, one has
\begin{eqnarray}\label{proof5}
 &&\no\hspace{-1.2cm}\frac{\delta [\omega_0 \mu E_1^1+(\omega_0\nu+\mu
\omega_1)E_1^0]}{\delta u}_{\mid_{\{\omega_0=u,
\omega_1=u\}}}\\\no&&\hspace{1cm}=u\Big[\frac{\partial \mu}{\partial
u} E^1_1+\frac{\partial (\mu+\nu)}{\partial u}
E^0_1\Big]+\frac{\delta (\varphi E^1_1+\phi E^0_1)}{\delta
u}_{\mid_{\{\omega_0=u, \omega_1=u\}}}
\\&&\hspace{1cm}=\widetilde{\lambda}_1 E_1^0+\widetilde{\lambda}_0
E_1^1,
\end{eqnarray}
where $\phi=\omega_0\nu(x,u)+\omega_1\mu(x,u)$ is a new dependent
variable.

Assume equation $E_1=0$ is approximately nonlinearly self-adjoint,
then we have the multiplier $\mu+\epsilon\nu=\varphi/u+\epsilon
(\phi-\varphi)/u$, with the help of (\ref{proof2}),(\ref{proof4})
and (\ref{proof5}), Eq.(\ref{proof3}) becomes
\begin{eqnarray}\label{proof6}
 &&\no\hspace{-0.5cm}\frac{\delta [\omega (\mu(x,u)+\epsilon\nu(x,u)) E_1]}{\delta
u}_{\mid_{\omega=u+\epsilon u}}\\\no&&\hspace{0.6cm}=u
\frac{\partial \mu}{\partial u} E^0_1+\frac{\delta (\varphi
E^0_1)}{\delta u}_{\mid_{\omega_0=u}}+\epsilon u\Big[\frac{\partial
\mu}{\partial u} E^1_1+\frac{\partial (\mu+\nu)}{\partial u}
E^0_1\Big]+\epsilon\frac{\delta (\varphi E^1_1+\phi E^0_1)}{\delta
u}_{\mid_{\{\omega_0=u, \omega_1=u\}}}\\\no&&\hspace{0.6cm}
=\Big[\lambda_0+\frac{\partial \varphi}{\partial
u}-\frac{\varphi}{u}\Big] E^0_1+\epsilon
\Big[\Big(\lambda_0+\frac{\partial \varphi}{\partial
u}-\frac{\varphi}{u}\Big) E^1_1+\Big(\lambda_1+\frac{\partial
\phi}{\partial u}-\frac{\phi}{u}\Big) E^0_1\Big]
\\\no&&\hspace{0.6cm}
=\widetilde{\lambda}_0E_1^0+\epsilon(\widetilde{\lambda}_1
E_1^0+\widetilde{\lambda}_0 E_1^1),
\end{eqnarray}
thus \begin{eqnarray}
\no\widetilde{\lambda}_0=\lambda_0+\frac{\partial \varphi}{\partial
u}-\frac{\varphi}{u},~~~
\widetilde{\lambda}_1=\lambda_1+\frac{\partial \phi}{\partial
u}-\frac{\phi}{u}.
\end{eqnarray}
Hence, equation $E_1=0$ multiplied by the multiplier
$\mu+\epsilon\nu$ is approximately strictly self-adjoint.

Conversely, let $E_1=0$ with
multiplier $\mu+\epsilon\nu$ is approximately strictly self-adjoint,
then taking $\varphi=u\mu,\phi=u(\mu+\nu)$, Eq.(\ref{proof2})
becomes
\begin{eqnarray}\label{proof7}
\no&&\frac{\delta (\varphi E_1^0)}{\delta u}=
\Big[\widetilde{\lambda}_0-u \frac{\partial \mu}{\partial u}\Big]
E_1^0, \\\no&&\frac{\delta (\varphi E_1^1+\phi E_1^0)}{\delta u}=
\Big[\widetilde{\lambda}_0-u\frac{\partial \mu}{\partial u}\Big]
E_1^1+\Big[\widetilde{\lambda}_1-u\frac{\partial (\mu+\nu)}{\partial
u} \Big]E^0_1.
\end{eqnarray}
Alternatively,
\begin{eqnarray}\label{proof8}
\no&&\frac{\delta (v E_1)}{\delta u}_{\mid_{v=\varphi+\epsilon
\phi}}=\Big[\widetilde{\lambda}_0-u \frac{\partial \mu}{\partial
u}\Big] E_1^0+\epsilon \Big[\widetilde{\lambda}_0-u\frac{\partial
\mu}{\partial u}\Big] E_1^1+\epsilon
\Big[\widetilde{\lambda}_1-u\frac{\partial (\mu+\nu)}{\partial u}
\Big]E_1^0,
\end{eqnarray}
then \begin{eqnarray} &&\no\lambda_0=\widetilde{\lambda}_0-u
\frac{\partial \mu}{\partial
u},~~~\lambda_1=\widetilde{\lambda}_1-u\frac{\partial
(\mu+\nu)}{\partial u}.
\end{eqnarray}
We conclude that equation $E_1=0$ is approximately nonlinearly
self-adjoint, thus complete the proof. $\blacksquare$

\textbf{Remarks 3.}

1. Theorem 3 holds for some special cases of
$\lambda_i,\widetilde{\lambda}_i \,(i=0,1)$. For example, if
$\lambda_1=\widetilde{\lambda}_1=\phi=0$, then one find that the
substitution is given by $v=\varphi$ and the multiplier becomes
$\mu=\varphi/u$,  which has the same results as unperturbed case
\cite{ib-2011}.

2. If the substitution for approximate strict self-adjointness is
adopted by $v=u$, i.e., $\omega_0=u,\omega_1=0$ in the proof of
Theorem 3, then we have the multiplier $\mu+\epsilon\nu$ is taken
the form $(\varphi+\epsilon\phi)/u$. Similarly, if the substitution
is $v= \epsilon u$,  we have the multiplier
$\mu+\epsilon\nu=\epsilon \phi/u$.

\section{Applications}

In this section, we apply approximate nonlinear self-adjointness to
construct approximate conservation laws of a class of perturbed
nonlinear wave equations
\begin{eqnarray}\label{wave}
u_{tt}-[F(u)u_x]_x+\epsilon u_t=0,~~F'(u)\neq 0,
\end{eqnarray}
where $F(u)$ is an arbitrary smooth function. Eq.(\ref{wave})
describes wave phenomena in shallow water, long radio engineering
lines and isentropic motion of a fluid in a pipe
etc.\cite{ba-1991,char}. The perturbing term $\epsilon u_t$ arises
in the presence of dissipation and the function $F(u)$ is defined by
the properties of the medium and the character of the dissipation.

Eq.(\ref{wave}) had been studied by means of the stated two
approximate symmetry methods and affluent approximate solutions were
obtained \cite{zhang1,ba-1991}. 
For Eq.(\ref{wave}), we take the following formal Lagrangian
\begin{eqnarray}\label{lag}
&& \mathcal {L}=v\left[u_{tt}-[F(u)u_x]_x+\epsilon u_{t}\right],
\end{eqnarray}
and work out the variational derivative of this formal Lagrangian to
obtain the system of two coupled equations
\begin{eqnarray}\label{as}
&&\no \frac{\delta \mathcal {L}}{\delta v}=u_{tt}-[F(u)u_x]_x+\epsilon u_t=0,\\
&&\frac{\delta \mathcal {L}}{\delta u}=v_{tt}-F(u)v_{xx}-\epsilon
v_t=0,
\end{eqnarray}
where the second equation is called the adjoint equation of
Eq.(\ref{wave}).

\subsection{Approximate nonlinear self-adjointness}

With the help of computable formula (\ref{forper}), we prove the
following proposition for Eq.(\ref{wave}).

 \textbf{Proposition 1.}
Eq.(\ref{wave}) is approximately  nonlinearly self-adjoint under the
substitution
\begin{eqnarray}\label{substitution}
v=(c_1t+c_2)x+c_3t+c_4+\epsilon\left[(\frac{1}{2}c_1t^2+c_5t+c_6)x+\frac{1}{2}c_3t^2+c_7t+c_8\right],
\end{eqnarray}
where $c_i(i=1,\dots,8)$ are arbitrary constants such that $v\neq0$.

 \emph{Proof.}
Assuming that $v=\varphi(x,t,u)+\epsilon \phi(x,t,u)$ and
substituting it into the adjoint equation, one obtains
\begin{eqnarray}\label{det}
&&\no\varphi_{tt}+2\varphi_{tu}u_t+\varphi_{uu}u_t^2+\varphi_uu_{tt}
+\epsilon(\phi_{tt}+2\phi_{tu}u_t+\phi_{uu}u_t^2+\phi_uu_{tt}-\varphi_t-\varphi_uu_t)\\\no
&&-F(u)[\varphi_{xx}+2\varphi_{xu}u_x+\varphi_{uu}u_x^2+\varphi_uu_{xx}
+\epsilon(\phi_{xx}+2\phi_{xu}u_x+\phi_{uu}u_x^2+\phi_uu_{xx})]\\
&&=(\lambda_0+\epsilon\lambda_1)[u_{tt}-F'(u)u_x^2-F(u)u_{xx}]+\epsilon
\lambda_0 u_t,
\end{eqnarray}
where we omit the second-order terms of $\epsilon$ in
Eq.(\ref{det}).

Comparing the coefficients for $u_{tt},u_{xx},u^2_x,u^2_t$ in both
sides, we obtain $\varphi_u=\phi_u=0$ and $\lambda_0=\lambda_1=0$,
then the above equation (\ref{det}) becomes
$\varphi_{tt}+\epsilon(\phi_{tt}-\varphi_t)-F(u)(\varphi_{xx}+\epsilon\phi_{xx})=0$
and yields
\begin{eqnarray}
&&\no\varphi_{tt}=0,~~\phi_{tt}-\varphi_t=0,~~\varphi_{xx}=0,~~\phi_{xx}=0,
\end{eqnarray}
which gives the solutions
$\varphi=(c_1t+c_2)x+c_3t+c_4,\phi=(\frac{1}{2}c_1t^2+c_5t+c_6)x+\frac{1}{2}c_3t^2+c_7t+c_8$.
This completes the proof. $\blacksquare$

Proposition 1 provides many choices for multipliers to make
Eq.(\ref{wave}) become approximately strictly self-adjoint. For
example, after assigning proper values to some parameters,  we have
$v= 1+\epsilon$. Multiplying it on the left side of Eq.(\ref{wave}),
by means of Theorem 3, we obtain $\mu(x,t)=1/u,\nu(x,t)=0$, then
Eq.(\ref{wave}) multiplied by it becomes
\begin{eqnarray}\label{wave1}
\frac{1}{u}\Big[u_{tt}-[F(u)u_x]_x+\epsilon u_t\Big]=0,
\end{eqnarray}
which is approximately strictly self-adjoint because, at this time,
$\mathcal {L}=v[u_{tt}-[F(u)u_x]_x+\epsilon u_t]/u$ and the adjoint
equation of Eq.(\ref{wave1}) is
\begin{eqnarray}
&&\no\frac{\delta \mathcal{L}}{\delta u}=
\frac{1}{u^3}\big[u^2v_{tt}-2uu_tv_t-2uvu_{tt}+2vu_t^2-\epsilon
u^2v_t\big]\\\no&&\hspace{1.1cm}+\frac{1}{u^3}[u
F'(u)-2F(u)]vu_x^2+\frac{1}{u^2}[2u_xv_x+2vu_{xx}-uv_{xx}]F(u)=0,
\end{eqnarray}
which becomes Eq.(\ref{wave1}) by substitution $v=u$. Alternatively,
one can adopt $\mu=\nu=1/u$ to make Eq.(\ref{wave}) to be
approximate strict self-adjointness.
\subsection{Approximate conservation laws}
Now we turn to construct approximate conservation laws of
Eq.(\ref{wave}). The first step of the approach is to perform
approximate symmetry classification of Eq.(\ref{wave}). The exact
symmetry of unperturbed equations
\begin{eqnarray}\label{unperturbed}
&&u_{tt}-[F(u)u_x]_x=0,
\end{eqnarray}
is well known \cite{wf-1981}. The maximal Lie algebra is generated
by a three-dimensional algebra and three special cases correspond to
four- or five-dimensional Lie algebra. The results are reduced to
those cases in Table 1 by the equivalence transformation
\begin{eqnarray}
&&\no\widetilde{x}=e_1x+e_2,~~\widetilde{t}=e_3t+e_4,~~\widetilde{u}=e_5u+e_6,
\end{eqnarray}
where $e_i,(i=1,\dots,6)$ with $e_1e_3e_5\neq0$ are arbitrary
constants.
\begin{center}
 {\bf Table 1}.  Lie algebras  of Eq.(\ref{unperturbed})
\end{center}
{\renewcommand{\arraystretch}{1.1}\begin{center}
\newcommand{\rb}[1]{\raisebox{3ex}[0pt]{#1}}
\begin{tabular}{p{3cm}p{11.5cm}}\hline
\multicolumn{1}{c}{$F(u)$}
 &\multicolumn{1}{c}{Symmetry Operators}\\\hline
 $\text{arbitrary}$ &$X_1=\partial_x,X_2=\partial_t,X_3=x\partial_x
+t\partial_t$\\
$e^{u}$ &$X_1,X_2,X_3,X_4=x\partial_x +2\partial_u$\\
$u^{\mu}(\mu\neq-4,-\frac{4}{3})$&$X_1,X_2,X_3,X_5=\mu x\partial_x +2u\partial_u$\\
$u^{-4}$ &$X_1,X_2,X_3,X_5,X_6=t^2\partial_t +tu\partial_u$\\
$u^{-\frac{4}{3}}$ &$X_1,X_2,X_3,X_5,X_7=x^2\partial_x
-3xu\partial_u$
\\\hline
 \end{tabular}
 \end{center}}

Table 2 gives approximate symmetries obtained by the approach of
Baikov et al. \cite{ba-1991}.
\begin{center} {\bf Table 2.} Approximate Lie algebras  of Eq.(\ref{wave})
\end{center}
{\renewcommand{\arraystretch}{1.2}\begin{center}
\begin{tabular}{p{2.9cm}p{11.8cm}}\hline
\multicolumn{1}{c}{$F(u)$}&\multicolumn{1}{c}{Approximate symmetry
operators}\\\hline
 arbitrary& $\widetilde{X}_1=\partial_x,\widetilde{X}_2=\partial_t,\widetilde{X}_3=\epsilon(x\partial_x+t\partial_t)$\\
$e^u$&
$\widetilde{X}_1,\widetilde{X}_2,\widetilde{X}_3,\widetilde{X}_4=x\partial_x+t\partial_t+\epsilon
t(\frac{t}{2}\partial_t-2\partial_u)$\\
 $u^{\mu}(\mu\neq-4,-\frac{4}{3})$& $\widetilde{X}_1,\widetilde{X}_2,\widetilde{X}_3,
 \widetilde{X}_5=\epsilon(x\partial_x+\frac{2}{\mu}u\partial_u),
  \widetilde{X}_6=t\partial_t-\frac{2}{\mu}u\partial_u
  +\epsilon \frac{\mu t}{\mu+4}(\frac{t}{2}\partial_t-\frac{2}{\mu}u\partial_u)$\\
   $u^{-4}$& $\widetilde{X}_1,\widetilde{X}_2,\widetilde{X}_3,\widetilde{X}_5,\widetilde{X}_6,
   \widetilde{X}_7=\epsilon(t^2\partial_t+tu\partial_u)$\\$u^{-4/3}$&
   $\widetilde{X}_1,\widetilde{X}_2,\widetilde{X}_3,\widetilde{X}_5,\widetilde{X}_6,
  \widetilde{X}_8=\epsilon(x^2\partial_x-3xu\partial_u)$\\\hline
 \end{tabular}
 \end{center}}

In what follows, we will use formula (\ref{formula}) in Theorem 1 to
obtain first-order approximate conservation laws of Eq.(\ref{wave})
\begin{eqnarray}
&&\no\left[D_t(C^1)+D_x(C^2)\right]_{|(\ref{wave})}= O(\epsilon^2),
\end{eqnarray}
by virtue of the approximate symmetries in Table 2.

Inserting the formal Lagrangian (\ref{lag}) into formula
(\ref{formula}), we obtain
\begin{eqnarray}\label{vector}
&&\no C^1=W(\epsilon v-v_t)+v D_t
(W),\\&&C^2=W\left(F(u)v_x-F'(u)u_x v\right)-D_x(W) F(u) v.
\end{eqnarray}

In particular, we investigate the following two cases to illustrate
the effectiveness of approximate nonlinear self-adjointness in
constructing approximate conservation laws, while other cases can be
done by formula (\ref{vector}) with similar procedure.

\textbf{ Example 1.} Now, we utilize operator $\widetilde{X}_4$ in
Table 2 to calculate the conserved vector. In this case,
$W=-2\epsilon t-x u_x-\big(t+\frac{1}{2}\epsilon t^2\big)u_t$ and
$F(u)=e^u$, then  the conserved vector (\ref{vector}) becomes
\begin{eqnarray}\label{cons11}
&&\no C^1= t e^{u}u_x v_x+x v_tu_x+ t u_tv_t+x
u_tv_x\\\no&&\hspace{1cm}+\frac{1}{2}\epsilon \left(t^2 u_tv_t+4 t
v_t+t^2 e^{u} u_x v_x-2t u_tv-2x u_xv-4v\right),
   \\\no&&C^2=-t e^{u} u_tv_x
   -x u_tv_t-te^{u} u_x v_t-xe^{u} u_x
   v_x\\&&\hspace{1cm}
   +\epsilon\big(x u_tv+t e^{u} u_xv-\frac{1}{2} t e^{u}
   (t u_x v_t+t u_tv_x+4
   v_x)\big),
\end{eqnarray}
 where $v$ is given by the substitution (\ref{substitution}).

 Specially, we take $v=x$, then (\ref{cons11}) becomes
\begin{eqnarray}
&&\no C^1_1=x u_t+t e^{u} u_x+\epsilon (\frac{1}{2} t^2 e^{u} u_x-
x^2 u_x- t x u_t),\\\no&&C^2_1=-t e^{u} u_t-x
e^{u}u_x+\epsilon(-\frac{1}{2} t^2 e^{u} u_t+x^2 u_t+t x e^{u} u_x-2
t e^{u}),
\end{eqnarray}
which make
\begin{eqnarray}
\no\Big[D_t(C^1_1)+D_x(C^2_1)\Big]_{|(\ref{wave})}= \epsilon ^2 x
\left(t u_t-2\right).
\end{eqnarray}

 \textbf{ Example 2.} Consider
$\widetilde{X}_6$ for $F(u)=u^{\mu}(\mu\neq-4,-\frac{4}{3})$. Here,
$W=-\frac{2}{\mu}u - \frac{2\epsilon
  tu}{\mu+4}-\big(t+\frac{\epsilon\mu
  t^2}{2(\mu+4)}\big)u_t$, then  the
conserved vector (\ref{vector}) becomes
\begin{eqnarray}
&&\no C^1= t u_x v_x u^\mu+ t u_t v_t- u_t v-\frac{2}{\mu} (u_t v-2
u v_t)\\\no&&\hspace{1cm}+\frac{\epsilon}{2 \mu (\mu+4)}\left(\mu^2
t^2 u_x v_x u^\mu+\mu^2 t^2u_t v_t-2\mu  (\mu+2)t u_t v- 8(\mu+2)u
v+4\mu tuv_t\right),\\\no&&C^2= u^\mu u_x v-tu^\mu u_x v_t- tu^\mu
v_x u_t+\frac{2}{\mu} \left(u_xv-v_x
u\right)u^\mu\\&&\hspace{1cm}-\frac{\epsilon t u^\mu }{2
(\mu+4)}\left(\mu t u_x v_t+ \mu t u_tv_x+4 uv_x-2 (\mu+2) u_x
v\right),
\end{eqnarray}
where $v$ is also given by the substitution (\ref{substitution}).

 In particular, choosing $v=x+\epsilon t$, we have
\begin{eqnarray}
&&\no C^1_2=tu^\mu u_x -(\frac{2}{\mu}+1) x
u_t\\\no&&\hspace{1cm}+\frac{\epsilon}{2 \mu (\mu+4)}\left(\mu^2 t^2
u^\mu u_x-2 t \left(\mu^2 x+2 \mu (x+1)+8\right) u_t-4 (2 \mu
x-\mu+4 x-4) u\right),
   \\\no&&C^2_2=x u_xu^\mu-t u_tu^\mu +\frac{2}{\mu}\left(x u_x- u\right)u^\mu
   \\\no&&\hspace{1cm}-\frac{\epsilon t u^\mu }{2 \mu (\mu+4)}\left(\mu^2 t u_t-2 \mu^2 xu_x-4 \mu (x+1)u_x-16 u_x+4 \mu u\right),
\end{eqnarray}
and
\begin{eqnarray}
&&\no\hspace{-0.2cm}\Big[D_t(C^1_2)+D_x(C^2_2)\Big]_{|(\ref{wave})}\\\no&&\hspace{0.6cm}=-\frac{\epsilon
^2}{2 \mu  (\mu +4)} \left(2\mu^2
 t u_t-\mu^2t(2x+\epsilon t)u_{t}+8\mu t u_t+4
\mu u(2 t \epsilon +x+1)+16u\right).
\end{eqnarray}
\section{Conclusion}
We provide a more detailed investigation of approximate nonlinear
self-adjointness and related properties, which extends the results
in both the unperturbed PDEs case and the perturbed ODEs case.
The results are applied to a class of perturbed
nonlinear wave equations and approximate conservation laws are
obtained.

\section*{Acknowledgments}
The author is grateful to Prof. N.H. Ibragimov for constructive
suggestions and helpful comments. I thank the anonymous referees for
their valuable advices and corrections to the paper.

\end{document}